\documentclass{llncs}
    \usepackage{graphicx}
    \usepackage{wrapfig}
\usepackage{amssymb}       %%% mathematical symbols
\usepackage[latin1]{inputenc}

    \usepackage[a4paper=true,  bookmarks=false, linkcolor=black, citecolor=black,
      urlcolor=black,colorlinks=true,pagecolor=black, breaklinks=true, bookmarksopen=false,
      pdftitle={Notations Around the World:Census and Exploitation},
      pdfauthor={Paul Libbrecht},
      pdfkeywords={mathematical notations,rendering,formulae,symbols,learning-content},
     ]{hyperref}

\begin{document}

\title{Notations Around the World:\\Census and Exploitation}
  \author{Paul Libbrecht
     \institute{DFKI GmbH and University of Saarland, Saarbr\"ucken, Germany}
     \email{paul@(activemath.org$|$dfki.de) }}
\titlerunning{Notations Around the World...}
\maketitle              % typeset the title of the contribution

\begin{abstract}

Mathematical notations around the world are diverse. Not as much as
  requiring computing machines' makers to adapt to each culture, but as
  much as to disorient a person landing on a web-page with
  a text in mathematics.

In order to understand better this diversity, we are building a
  census of notations: it should allow any content creator or mathematician to grasp
  which mathematical notation is used in which language and culture.
  The census is built collaboratively, collected in pages with a given
  semantic and presenting observations of the widespread notations
  being used in existing materials by a graphical extract. We contend
  that our approach should dissipate the fallacies found here and there
  about the notations in ``other cultures" so that a better
  understanding of the cultures can be realized.

The exploitation of the census in the math-bridge project is also
  presented: this project aims at taking learners ``where they are in
  their math-knowledge" and bring them to a level ready to start
  engineering studies. The census serves as definitive reference for the
  transformation elements that generate the rendering of formulæ in
  web-browsers.\footnote{The final publication of this paper is available at \url{http://www.springerlink.com/}.}
\end{abstract}

\section{Introduction}
% --------------------------------------

This paper reminds the great variation in the notations used in mathematical texts in many cultures: in order to obtain knowledge of this diversity, we propose to realize a census of the mathematical notations made of concrete observations from multiple texts. This census allows us to reach verifiable statements about the notations in wide use.

Misconceptions about widely-used notations are quite easy to reach: indeed, it is almost impossible to obtain persons that master the mathematical notations of several languages. Thus, one can find in several texts about mathematical notations the fact that the binomial coefficient in Russian is written as ${\mathrm C}_k^n$ whereas the binomial coefficient in French is written as ${\mathrm C}_n^k$, i.e. they are opposite of each other. This is the case of \cite{MathML}, \cite{OMDoc-1.2}, and \cite{LibbrechtAndresGu-SmartPaste-MathUI09}. As the authors could observe and proof in many russian sources, this turned out to be false: the binomial coefficient in Russian is written the same as in French. Tracking down the source of this confusion turned out to be impossible with the inability to verify the origin of such a claim and with, even, easy assertions about the pre-industrial era where such a notation was carried between France and Russia.

The census we propose should avoid such misconceptions by the maintenance of web-pages that link to textbooks that are commonly used hence should allow to dissipate any doubt about notations in wide usage in different cultures even if it is not well known.

We also aim at exploiting the census for the presentation system of ActiveMath, a web-based environment that renders mathematical formulæ from a semantic source into a presentation suitable for the learner. The Math-Bridge project's objective is to start where the learners start: in their own mathematical culture (the background can be quite different as described in~\cite{MelisEtalCultureJournal2009} and~\cite{intergeo-D2.5}, and bridge the gap to reach university-entry-level mathematics. Therefore, a census of college-level text-books' notations for the languages of Math-Bridge is useful: English, German, French, Spanish, Finnish and Hungarian.

In this paper we explain how notations are exploited to create the necessary information that instructs the ways of rendering of the presentation system of ActiveMath: it attaches an OpenMath {\it prototype}, within a given {\it user-context}, to a MathML-presentation expression. One of the conclusions appearing here is that the user-contexts can be quite multiple. We explain the architectural choices that allow these user-contexts to be fine-grained but still perform high-speed rendering of the pages.

\subsection{Outline}
% --------------------------------------

The paper first introduces related works, then proposes a broad classification of the notational diversity found around the world. The research litterature that speaks about the notational diversity is covered. The paper then presents the ingredients and principles of the notation census and indicates some typical examples. The exploitation within the ActiveMath environment is described. An outlook concludes the paper.

\section{Related Works}
% --------------------------------------

One of the most notable and comprehensive census of mathematical notations was done by Florian Cajori in~\cite{Cajori-Notations} in 1928 who concentrated on the English language. Encyclopediae of mathematics could also claim to a very broad coverage but all of them are monolingual with the exception of the WikiPedia initiative, created by a great amount of contributors; not surprisingly, even the notations of the English WikiPedia is not fully consistant: one finds for example the binomial coefficient written one way or another depending on the page. 

The adaptation needed so that a software can interact with its users dependending on the cultures of the world have been investigated by such works as~\cite{Aaron-Marcus-Global-UID} often based on the influential work of Hofstede~\cite{Hofstede2001}. However, such studies rarely went as detailed as to indicate the variability of mathematical formulæ.

Within the mathematical knowledge communities several papers quote the need to respect mathematical diversity, e.g.~\cite{Manzooretal-Notations-MKM-2005,mueller-et-al-documents-notations-interfaces,Smirnova-TeX-with-Notations,OpenMath-Utility} but the examples used there are too restricted to be representative. The confusion indicated above about the notation for binomial coefficient has crept into~\cite{MathML} and been cited by several such papers as example.

We have, in~\cite{MelisEtalCultureJournal2009} presented some of the diversity we have met and some approaches to a solution. However this paper aimed at presenting the issues rather than aim at comprehensiveness. Therefore, we describe here the approach towards reaching the coverage needed for an informed usage of mathematical notations.

\section{Notational Diversity Accross Cultures}
% --------------------------------------

Mathematics is widely viewed as a universal ``methodology of reasoning'' and a ``language'' common to all humans, truly ``objective'', and thus independent of subjective beliefs and cultural preformation. In particular, statements written in the mathematical symbol language, i.e. in terms of ``formulæ'', should thus be understandable in principle by everyone. In a general sense this is indeed true, and it is exactly this fact that makes mathematics a widely applicable tool for (almost) all sorts of problems: The mathematical language provides the possibility to express abstract concepts and logic reasoning in an unambigous way that can be understood independent of the reader's language and culture. If we say that a mathematical theorem (for example Fermat's last theorem) has been {\it proven}, we mean that a certain mathematical truth has been acertained at an objective level: It has been shown to hold {\it as such}, without reference to a particular language or culture.

However, when {\it practicing} mathematics (no matter whether the application of pre-established methods or genuine mathematical creativity are concerned, whether easy or difficult), language and culture dependent issues do play a role at several levels. Mathematical thought has evolved in history in different cultural contexts (beginning with several hundred years B.C.), each developing its own methods by which mathematical ideas can be expressed and operated with. Although the development of modern science and a world-wide scientific communication within the last few centuries lead to an extensive unification and ``internationalization'' of mathematical conventions, a lot of variations remained, so that we may talk about ``mathematical cultures'', which sometimes (but not always) are associated with ``cultures'' in the sense of countries or languages. In particular, differences exist even between regions as close to each other as the European countries.

To begin with, mathematical statements are usually embedded in phrases of ``ordinary language'', which may be viewed as ``everyday language'' used in a more or less rigorous way. For example, the statement

$$\textsf{there exists a natural number}\,\, n\,\, \textsf{such that}\,\, n^2=4$$

contains the verbal elements ``\textsf{there exists}'', ``\textsf{natural number}'' and ``\textsf{such that}''. It is {\it in principle} possible to completely omit verbal elements of this type. For example, statement above could be re-expressed as $\exists n\in {\mathbb N}:n^2 =4\,.$

However, in most cases a rigorous elimination of all ``words between the symbols'' would be tremendously impractical, in particular in texts addressing mathematics students (who are just about to {\it learn} mathematics) and non-specialists (interested to apply mathematical techniques for various purposes). As a consequence, mathematical texts in practice rely on written {\it words} at least as much as on written {\it mathematical symbols}, thus suffering from all sorts of variations in meaning and difficulties of translation.

One might expect now that at least the mathematical {\it symbols} provide something like a well-defined and culturally independent language (or at least vocabulary). However, mathematical concepts are not only used in the contemporary international mathematical research communication -- which indeed uses a largely standardized symbolic language and may thus be considered as a ''mathematical culture'' in its own -- but also in local contexts (e.g. technical or economic) or just in every-day life. All these fields of usage of mathematical concepts constitute ``cultures'' whose customs differ in many respects.

\subsection{Differences in Decimal Numbers}
% --------------------------------------

The most basic example is how {\it numbers} are written. Whereas in many countries (as well as in the international communication) the number ``twelve and a half'' is written in decimal representation as $12.5$ some mathematical cultures would use  $12,5$ instead. In the first form {\it decimal separator} is indicated by a period or point (we thus say in English language ``twelve point five''), whereas in the second a comma is used instead (consequently, in German language one calls this number ``zw\"olf Komma f\"unf'' and in the French language, one calls this number ``douze virgule cinq''). Having evolved historically, notational differences of this type have even been standardized in local contexts. For example, the DIN (Deutsch Industrie-Norm -- German industrial norm) specifies the comma as used in  the second form as the ``official'' decimal separator. Students in German schools are thus acquainted with the notation $12,5$ rather than $12.5$. Note however that the {\it meaning} of the period in the first and the comma in the second are precisely the same in both cases! In technical terms, these two differing notations constitute {\it semantically equivalent concepts}.

Consequently, a general rule such as $\textsf{the period means the same as the comma}$ spelled out once and for all, could be admissible in a mathematical text that uses the period notation but is otherwise written in German. However, such rules are most often simplistic and fail because the other character (the comma or period) is commonly used as thousands separator. Thus, on the web, a german user may well encounter and be confused by $1.001$ which could mean $1+\frac{1}{1000}$ or $10^3 +1$. So as to attempt clarification, the notation census thus contains a page about decimal numbers: \url{http://wiki.math-bridge.org/display/ntns/Decimal+Numbers}. A comprehensive set of number patterns is assembled at \url{http://unicode.org/repos/cldr-tmp/trunk/diff/by_type/number.pattern.html} albeit quite technical.

\subsection{Difference Because of names Differences}
% --------------------------------------

Mathematical concepts and operations are often {\it named} by words or phrases and these words are sometimes seen in the notations. For example the ``sine'' function (German: ``Sinus'', Spanish: ``seno''). Just by their use, verbal expressions of this type have become part of the (scientific) language of the speaker or reader. Which word or phrase is used to denote a concept or an operation is to some extent a matter of (official or informal) standardization, but it might as well be a matter of {\it cultural tradition}. As the above examples show, the particular word or phrase used to denote {\it one} concept will be different when expressed in different languages. Most mathematicians in the world would use the symbol ``$\sin$'' as an abbreviation of ``sine'' (``Sinus''), so that we may find a formula such as $\sin (\pi )=0$ in a mathematical textbook. However, in Spanish language (or ``culture'') of mathematics the abbreviation for this concept is ``${\rm sen}$''. A textbook written in Spanish language will thus rather contain the formula ${\rm sen}(\pi )=0$. Of course Spanish mathematicians are aware of this difference, and when addressing an international audience, they would use ``$\sin$'' instead of ``${\rm sen}$'', but it is a fact that Spanish pupils and students are familiar with ``${\rm sen}$'' rather than ``$\sin $''.

Similarly, in Arithmetics, even at an elementary level such as in secondary school, one knows the concept of the ``greatest common divisor'' of two integer numbers. In many languages the symbol used to denote this concept just consist of the initial letters of the corresponding verbal phrase. In English it is thus written as ``$\gcd$''. In the German tradition the same concept is called ``gr\"o{\ss}ter gemeinsamer Teiler'' and abbreviated  as ``${\rm ggT}$''. In Dutch, the corresponding phase ``grootste gemene deler'' is abbreviated as ``${\rm ggd}$''. In French the corresponding phrase is generally called plus grand commun diviseur and is written``${\rm pgcd}$'' though modernists tend to change it to ${\rm pgdc}$ because the formulation {\it commun diviseur} is archaic. See the notation census page about it: \url{http://wiki.math-bridge.org/display/ntns/gcd}.

\subsection{Differences to Avoid Confusion}
% --------------------------------------

Sometimes variations in the mathematical notation are dictated by the wish of clarity in the given written work: for example authors prefer to write the ``binomial coefficients'' in the French way in order to avoid possible confusion with the 2D-vectors: what is denoted by $\left( {}_3^5 \right)$ in most mathematical cultures tends to appear in French and Russian textbooks in the form  $\mbox{C}_5^3$ instead (see the census page about the binomial coefficient: \url{http://wiki.math-bridge.org/display/ntns/binomial-coefficient}).

\subsection{Same Notations, Different Concepts}
% --------------------------------------

We have stated before that in general {\it one} mathematical concept may be represented by different cultures in different ways. However, in some cases, a particular name or phrase -- that may easily be translated between the languages -- denotes {\it different} concepts. The most prominent example is the notion of ``natural numbers''. Originally, it meant set of integer numbers greater or equal to $1$, i.e. $ 1,2,3,4,5\dots $ and was denoted by the symbol ${\mathbb N}$. However, for practical reasons, it would be better to include the zero, hence to define the set of natural numbers by $ 0,1,2,3,4,5\dots $.

Meanwhile, several mathematical cultures have followed this idea, while sticking to the original symbol ${\mathbb N}$. In order to denote the set of numbers without zero, one then usually writes something like ${\mathbb N}^+$ or ${\mathbb N}^*$. In the tradition in which the zero is not included, one may write ${\mathbb N}^0$ or ${\mathbb N}_0$ in order to denote the set with zero. As a consequence, when the symbol ${\mathbb N}$ (or the name ``natural numbers'') appears in a mathematical text, it could mean the integers with or without zero, depending on the mathematical culture it originates from.\footnote{In a {\it good} textbook this point is of course explicitly clarified}

The problem is also pointed out in Eric Weissteins {\it MathWorld} pages: ``Regrettably, there seems to be no general agreement about whether to include $0$ in the set of natural numbers'' \cite{Mathworld-Natural-Numbers}. {\it Wikipedia} also reflects this ambiguity in most pages on the subject of natural numbers.

It should be added that the web in general pulls the user out of a single textbook into easy jumps between many content sources, with varying degree of reliability. When {\it automated formula rendering} is concerned, ambiguities of this type must be taken into account in order to avoid confusion and let the user recognize easily which variant it is.

\section{The Notation Census}
% --------------------------------------

Because mathematical notations are very diverse and their context of occurrence is just as diverse, we propose to establish a {\bf census of mathematical notations}. That census should list all available mathematical notations that are widely spread around the world in a way that enables mathematics readers to see the mathematical notations used in the multiple cultural contexts even though they do not understand the language of the documents.

The census should be {\bf visual} because mathematical notations are a graphical artifact and their rendering in web-browsers should succeed in all situations: this requirement prevents the usage of elaborate display technologies of mathematical formul{\ae} so as to ensure the certainty of rendering the notation of that cultural context.

The census should be {\bf traceable} so that one can recognize who has written each part and {\bf commentable} so that its quality can be steadily improved; normal web-authoring practices similarly to those of Wikipedia or others apply here, together with the public visibility of any editing action.

The census should be displaying {\bf widely used notations} by relying on extracts of mathematical texts that are widely used themselves; this is fundamental so as to achieve usefulness of the census.

Finally, the census should be {\bf verifiable} because the misunderstanding among the cultures can be critically high: any interested reader should be able to find in just a few clicks who are the authors making a claim that a notation is widely used and in which cultural context it is used with the trustability of the source of notation serving as discussable reference point.

\begin{figure}\begin{center}\includegraphics[width=10cm]{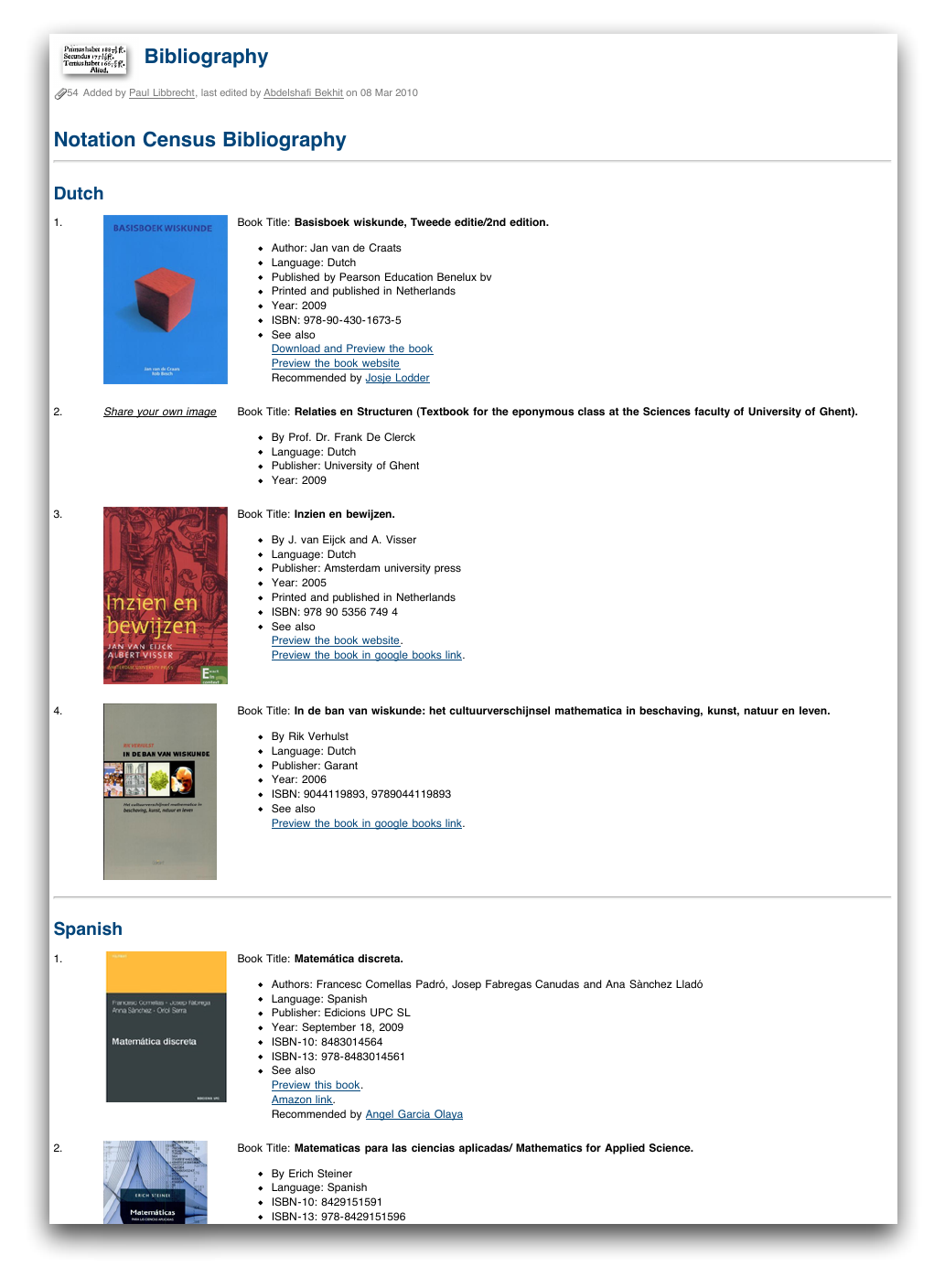}\caption{An extract of the bibliography}\end{center}\end{figure}

\subsection{Ingredients}
% --------------------------------------

We realize the notation census  in a {\bf public wiki} which is made of the following core ingredients:

\begin{itemize}

\item {\bf sources} are listed in a bibliography-like approach: on a single page containing all sources and a name of the cultural context they are used in; a link to a publisher page and link to possible web-download of the (partial) content. A source reference would be the entry point to judge the {\it relevance} of the notations with any given cultural context, something a person living in that context can do.

\item notations are grouped per page, {\bf one page per semantic}, each grouped in content-dictionaries following the semantic classification of, at least, the official OpenMath content dictionaries. Each semantic is linked to the OpenMath dictionary, the Wikipedia entry, and the entry in the Wolfram Encyclopedia~\cite{MathWorld} if possible.

\item notations are listed there with a small informal description, links to the content-dictionary entries,  and a series of {\bf observations} made of a text containing the observed notation, the name of the symbol in that context, a pointer to the element of the bibliography containing it, and a {\bf graphics copy} of the relevant bit indicated with a page number or other internal{\bf  reference} allowing a reader to find the used notation fast. This allows a predictable display, relying on graphical copies (scans or extracts of electronic versions of the book rendered in the browser as PNG or JPEG images). The name of the symbol in this culture is requested and, if possible, a character-based reproduction of the symbols' elements based on the Unicode character set.

\end{itemize}

The observations and sources both are given for a {\bf cultural context} whose definition is unprecise. It is generally accepted to be at least made of a human language but it often goes beyond it including traditions, domain of study, and the mere practical conventions in communities of practice. For example the French notation for binomial coefficient, with the big C, is recognized to be widespread at school levels, as can be seen in \url{http://wiki.math-bridge.org/display/ntns/binomial-coefficient} but several combinatorics researchers agree that the vector notation is preferable even in French~\cite{Bergeron-private-comm}. A more widespread example is that of the square root of $-1$ which is written mostly with the letter $i$ except in electrical engineering where the letter $i$ is too close to that of electrical current hence the square root of unity is written $j$; the list of different observed notations is presented on \url{http://wiki.math-bridge.org/display/ntns/nums1_i}. For this reason, the census speaks about the {\it cultural context} which can, typically,  be also defined by major texts.

Conformance is not, yet, strictly enforced for each contribution to the notation census: a check-list of ingredients before contributing an observation is provided in the {\it manifest}. We believe the requirements above, except maybe for a web-access to the text's content which is mentioned optional, are minimal.

\begin{figure}\begin{center}\includegraphics[width=10cm]{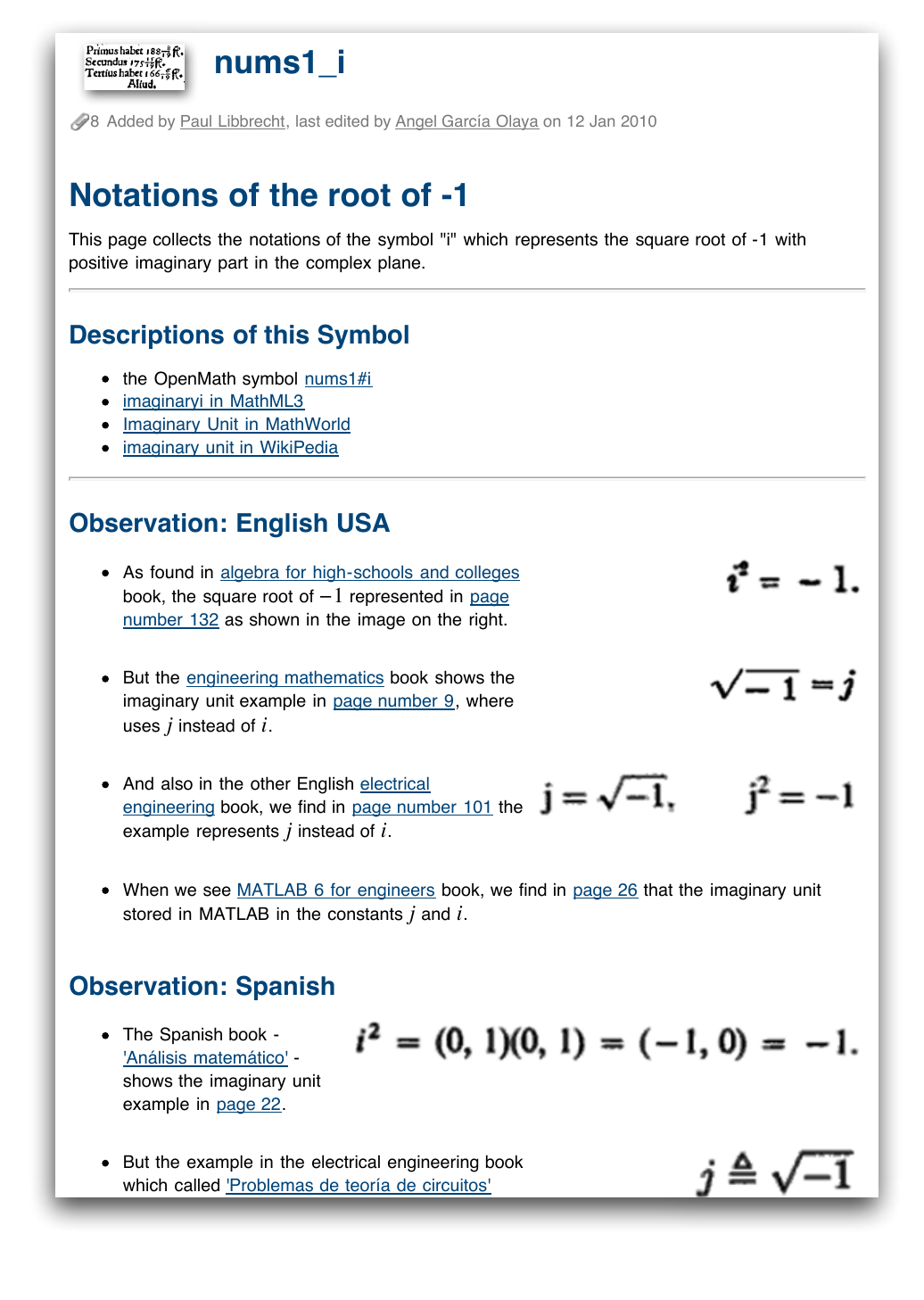}\caption{A page of observation: about the complex i.}\end{center}\end{figure}

\subsection{Evolution}
% --------------------------------------

The notation census can be reached at:

\begin{center}\url{http://wiki.math-bridge.org/display/ntns}\end{center}

It has started in October 2009.

The explanations are currently provided in the {\it notation census manifest} at \url{http://wiki.math-bridge.org/display/ntns/Notation-Census-Manifest}. This text intends to be the one stop information source about the census guiding ideas and principles.

We expect to populate the notation census gradually with the following contributions: first the MathML CD-group, then all ActiveMath available content, then the full Math-Bridge content.

The first objective of the notation census is to serve the math-bridge project which we describe below: the wiki is a good communication means between practicing mathematics educators and the technical team {\it encoding} the notations.

Beyond this project, it should support the ActiveMath project's development which is ongoing since more than 10 years. It should also support other initiatives of collecting notations for the rendering of semantic-mathematical-terms such as~\cite{mueller-et-al-documents-notations-interfaces}.

Finally, thanks to the connections to the sources, this census should also be one of the reliable places of information for a curious user wishing to be informed of the various ways to note mathematical terms in different cultures.

For each of these usages completeness is desirable but is not a requirement: the more coverage the better, but even a partial coverage can find its usages.

\section{Notations for ActiveMath}
% --------------------------------------

In this section, we turn to one of the exploitations: how the notation census will be exploited for the purposes of the Math-Bridge project and how the ActiveMath platform can be endowed to honour the cultural contexts by presenting mathematical formul{\ae} in a customizable way.

Math-Bridge intends to offer bridging-courses allowing learners to bridge mathematics knowledge gaps, taking them with their pre-existing knowledge, hence by the notations they already know, to a level ready to start University. Hence it aims to use the right notations in the cultural contexts the learner probably has learned in, that is late-school-level mathematics for Dutch, English, French, Finnish, German, Hungarian, and Spanish classes.

Math-Bridge intends to use the rich formul{\ae} rendering of the ActiveMath platform, coupled with other personalization features of this web-platform. This platform renders mathematical formu{\ae} from their OpenMath semantic source enabling a rich set of services attached to the rendering. These include the transfer to computer algebra systems, exercise inputs, plotter or the searchability...

ActiveMath presents mathematical formul{\ae} using a few widely supported display technologies on the web: HTML completed with CSS, XHTML with formul{\ae} in MathML, or PDF created by {\tt pdflatex}. The transformation from the OpenMath encoding to the rendering formats is done through several stages that are explained in~\cite{Ullrichetal-Presentation-ICALT04}. Two aspects are worth mentioning here:

\paragraph{Authorable Notations} ActiveMath rendering to HTML, XHTML, or TeX is done through XSLT and Velocity; most of the mathematical formul{\ae} conversion is done through {\tt symbolpresentation} elements which collect {\tt notation} elements each of which is a pair of an OpenMath expression and its associated rendering in MathML, see~\cite{Manzooretal-Notations-MKM-2005}. Each {\tt notation} element is annotated with a context of use: a notation is used when the prototype matches as well as the context.

\paragraph{Wealth of Configurable Contexts} As we have described above, the {\it cultures} in which mathematical notations are defined, are somewhat fuzzily defined.  Therefore ActiveMath allows the notations' cultures to be written in the following {\it contexts}:

\begin{itemize}
\item most importantly, a notation can be associated to a {\bf language}. At XSLT time, the notation of the right language is chosen.
\item a notation can be associated to an {\bf output format} (this is most important to care for imperfections of individual formats. At XSLT time, the notation for the right format is chosen.
\item a notation can be associated to an {\bf educational level}, indeed, we have often been requested that students at University see some operators one way while others in another way. This is realized by the notation generating the velocity code which is evaluated when this user-information is ready, at each delivery.
\item finally, a notation can be associated to a {\bf collection,} that is, to the collection of the book the item is in. This allows an author re-using a collection of someone else within his book, to ensure that formulæ in this book are used consistently, for his items or the one of others. This also evaluated at each delivery.
\end{itemize}

Within the Math-Bridge project, where a broad sharing of learning items is expected, these contexts are of fundamental importance to allow consistent mathematical notations within the context of each learning experience. Indeed, the most common review note we have received is about the notations whereas  these notations were fully checked for the LeActiveMath project by math educators that use it in their learning in German, Spanish, and English.

Based on the notation census, the content enrichment and translation processes will {\it encode} the necessary {\tt symbolpresentation} and {\tt notation} elements. This will be done as part of the encoding process and will be integrated in the quality proofing process.

\section{Conclusion}
% --------------------------------------

This paper has sketched the issues encountered to meet proper knowledge about mathematical notations in the multiple cultures and the great diversity of notations that are widely used in, at least, the secondary education mathematics texts. To address this lack of knowledge, we have proposed a notation census, in the form of a collaborative wiki-based effort collecting referenced and hyperlinked observations of the usage of widely used notations.

The census population has started in October 2009 and now contains about 130 observations. The duration to write a page collecting observations in the available sources is about 30 minutes with variations we have observed due to: the digital availability of the observations, the understanding by the encoder of the mathematical concept behind it, the searchability of the sources (a quality often missing in PDF books in Arabic we have been handling),  the availability of glyphs to search for (e.g. search for {\it gcd }is a lot easier than search for a matrix-transpose).

The wiki technology used here is quite basic and generic. Together with the limited requirement for structure, the potential for public contributions, without large technical competencies, is large. This lack, however, means a limited set of services: for example, it is not possible, yet, to list all the observations coming from one source or in a given language. Some of these can be recovered by an enriched tool set we intend to work on (sections in a wiki page are quite commonly automatically detected). Similarly the searchability is quite limited: the name of the symbol is the best key, provided it is known, but no {\it graphical search} can be done (except browsing through all); we believe the latter is a research challenge.

We intend to build on the census to help into creating links to web pages that speak about the mathematical symbols around the OpenMath content-dictionaries' web-pages in \url{http://www.openmath.org/}. The visual nature of the census is probably one of the efficient hooks for a mathematician's eyes which can enable him to identify that a given semantic is matching what he expects.

Finally, the collaborative nature of the census web system opens the long-term possibility of contributions of a broad body of contributors, similarly to that obtained by the Unicode consortium's Common Locale Data Repository as can be seen at~\url{http://cldr.unicode.org/index/charts}. Indeed, shortly following the public announce of this repository, in December 2009, spontaneous contributors appeared.

\section{Acknowledgements}
% --------------------------------------

The author wishes to thank Franz Embacher for fruitful input on how to present this work and Abdelshafi Bekhit for his numerous contributions to the notation census.

This work is partially funded by the European Commission under the eContentPlus programme in the Math-Bridge project (ECP-2008-EDU-428046).

\bibliographystyle{alpha}
%\bibliography{bibs/others,bibs/am,bibs/omega,bibs/learning_sys,local-bibs}

\newcommand{\etalchar}[1]{$^{#1}$}

\end{document}